\documentclass[12pt]{article}

\usepackage{scicite}

\usepackage{times}
\usepackage{graphicx}
\usepackage[dvipsnames]{xcolor}
\usepackage{ulem} 
\usepackage{hyperref}

\usepackage{lineno}

\newenvironment{sciabstract}{%
\begin{quote} \bf}
{\end{quote}}

\title{AutoPhaseNN: Unsupervised Physics-aware Deep Learning of 3D Nanoscale Bragg Coherent Diffraction Imaging} 

\author
{Yudong Yao,$^{1\ast}$ Henry Chan,$^{2}$ Subramanian Sankaranarayanan,$^{2}$ \\
Prasanna Balaprakash,$^{3}$ Ross J. Harder,$^{1\dagger}$ Mathew J. Cherukara$^{1\ddagger}$   \\
\\
\normalsize{$^{1}$Advanced Photon Source, Argonne National Laboratory, Lemont, IL 60439, USA}\\
\normalsize{$^{2}$Center for Nanoscale Materials, Argonne National Laboratory, Lemont, IL 60439, USA}\\
\normalsize{$^{3}$Mathematics and Computer Science, Argonne National Laboratory, Lemont, IL 60439, USA}\\
\\
\normalsize{$^\ast \dagger \ddagger$ To whom correspondence should be addressed;}\\ \normalsize{E-mail:  yudongyao@anl.gov, rharder@anl.gov, mcherukara@anl.gov}
}

\date{}
\begin{document} 


\baselineskip24pt


\maketitle 


\begin{sciabstract}
The problem of phase retrieval, or the algorithmic recovery of lost phase information from measured intensity alone, underlies various imaging methods from astronomy to nanoscale imaging. Traditional methods of phase retrieval are iterative in nature, and are therefore computationally expensive and time-consuming. More recently, deep learning (DL) models have been developed to either provide learned priors to iterative phase retrieval or in some cases completely replace phase retrieval with networks that learn to recover the lost phase information from measured intensity alone. However, such models require vast amounts of labeled data, which can only be obtained through simulation or performing computationally prohibitive phase retrieval on experimental datasets. Using a 3D nanoscale X-ray imaging modality (Bragg Coherent Diffraction Imaging or BCDI) as a representative technique, we demonstrate AutoPhaseNN, a DL-based approach which learns to solve the phase problem without labeled data. By incorporating the physics of the imaging technique into the DL model during training, AutoPhaseNN learns to invert 3D BCDI data from reciprocal space to real space in a single shot without ever being shown real space images. Once trained, AutoPhaseNN can be effectively used in the 3D BCDI data inversion about one hundred times faster than traditional iterative phase retrieval methods while providing comparable image quality.
\end{sciabstract}


\section*{Introduction}

The problem of phase retrieval is a central problem in many imaging techniques including X-ray Bragg coherent diffraction imaging (BCDI) and ptychography\cite{Miao2015}, electron ptychography\cite{Jiang2018ElectronResolution}, Lorentz transmission electron microscopy (LTEM)\cite{Phatak2016RecentMicroscopy}, super-resolution optical imaging\cite{Szameit2012Sparsity-basedImaging}, and astronomy\cite{Dean2006PhaseTelescope}. Phase retrieval is the algorithmic process of recovering phases from measured scattered intensities alone. In BCDI for example, a nanocrystalline sample is illuminated with a coherent X-ray beam from a synchrotron source or X-ray Free Electron Lasers (XFEL) and the scattered intensities are measured in the far-field at a Bragg peak. The measured intensities represent the modulus of the complex Fourier Transform (FT) of the object, but the phase of the wave is lost. Hence, the 3D image of the object cannot be recovered from a simple inverse FT and we must resort to phase retrieval algorithms that can recover this lost phase information to recover an image of the object. Additionally, when measured at a Bragg peak, the phase is influenced by the local strain within the crystal. Consequently, in addition to being a fundamental requirement to recovering the object's 3D structure, phase recovery also provides a 3D map of the strain state within the crystal, encoded as a phase of the complex image. This capability of BCDI to provide nanoscale structural information as well as picometer sensitivity to strain has had profound implications for the materials science, chemistry and solid-state physics communities. Examples include defect dynamics in battery electrodes\cite{Ulvestad2015}, in-situ catalysis\cite{Kim2018ActiveNanocrystals,kang2020time}, photon transport\cite{Cherukara2017a,Cherukara2017UltrafastNanostructures,Clark2013}, phase transformation\cite{Clark2015,Clark2015c,Ulvestad2015c}, and plastic deformation\cite{Hofmann20173DNano-crystals,Cherukara2018Three-dimensionalLoading,Yang2013}. 

More broadly, while coherent imaging techniques have grown to become an integral part of electron and X-ray materials characterization\cite{Jiang2018ElectronResolution,Phatak2016RecentMicroscopy,Pfeiffer2018X-rayPtychography}, their dependence on iterative phase retrieval to recover sample images prevents real-time feedback, which is particularly crippling for \textit{in-situ} and \textit{operando} experiments. Iterative phase retrieval typically requires thousands of iterations and often multiple starts to arrive at a robust solution, often taking longer than a single dataset acquisition time. 

Neural network (NN) models have been developed to rapidly solve inverse problems across a variety of disciplines including magnetic resonance imaging (MRI)\cite{Zhu2018ImageLearning}, image denoising\cite{Burger2012ImageBM3D,Lehtinen2018Noise2Noise:Data}, super-resolution\cite{Dong2016ImageNetworks,Ledig2017Photo-realisticNetwork,Lim2017EnhancedSuper-resolution}, etc. Specific to the problem of phase retrieval, neural networks have been trained to learn to retrieve phases in holographic imaging\cite{Rivenson2018PhaseNetworks}, lensless computational imaging\cite{Sinha2017LenslessLearning}, X-ray ptychography\cite{Cherukara2020AI-enabledImaging,Wengrowicz2020DeepPtychography,Guan2019PtychoNet:Ptychography}, Fourier ptychography\cite{Nguyen2018DeepMicroscopy} and in BCDI\cite{Cherukara2018Real-timeNetworks,Wu2021ComplexNetworks,Harder2021DeepImaging,chan2021rapid,scheinker2020adaptive}. Each of models in the aforementioned papers have been trained in a supervised manner, that is, training of the network uses pairs or triplets of images, including the experimentally acquired (or forward simulated) data and the known sample's complex image (typically amplitude and phase). NN training is then achieved by optimizing the weights of the network so that the network learns to output the sample image for a given measured data (the input diffraction pattern in the BCDI case). While this approach has been shown to provide speed, as well as reduced reciprocal space sampling requirements, in comparison to iterative phase retrieval, the approach is limited by the need for a large volume of labeled simulated data with corresponding images or untenable quantities of experimental data which has been inverted through traditional iterative phase retrieval. Both types of training data suffer from limitations. Simulated data is very often a poor substitute for real data, for example, it is hard to generate training data that is sufficiently diverse or is well representative of the experimental data, and it is usually free of experimental artefacts. On the other hand, performing phase retrieval on experimental data before training is a computationally and manually intensive task due to the number of hyperparameters that need to be adjusted for successful convergence. Recently, a new type of model, PhaseGAN, which is a generative adversarial network, has been trained to perform phase retrieval without pairing of inputs and outputs\cite{zhang2021phasegan}. But this approach also requires the availability of ground truth data, the subtle difference is that the inputs to the network and desired outputs do not need to be paired. Another recent study included a physics-informed neural network (PINN) which was trained in an unsupervised manner to recover the XFEL pulses in the time domain from the low-resolution measurements in both the time and frequency domains. However, the authors found PINN not as accurate as the network trained with supervised learning due to the lack of strong prior given by the labeled data \cite{ratner2021recovering}. 

In this work, we demonstrate the application of AutoPhaseNN, a physics-aware unsupervised deep convolutional neural network (CNN) that learns to solve the phase problem \textit{without ever being shown real space images of the sample amplitude or phase}. By incorporating the physics of the X-ray scattering into the network design and training, AutoPhaseNN learns to predict both the amplitude and phase of the sample given the measured diffraction intensity alone. Additionally, unlike previous deep learning models, AutoPhaseNN does not need the ground truth images of sample's amplitude and phase at any point, either in training or in deployment. Once trained, the physical model is discarded and only the CNN portion is needed which has learned the data inversion from reciprocal space to real space and is $\sim$100 times faster than the iterative phase retrieval with comparable image quality. Furthermore, we show that by using AutoPhaseNN's prediction as the learned prior to iterative phase retrieval, we can achieve consistently higher image quality, than neural network prediction alone, at 10 times faster speed than iterative phase retrieval alone. We demonstrate the application of AutoPhaseNN to 3D BCDI, motivated by the importance of the technique to materials characterization, especially in the light of upgrades to synchrotron sources and XFELs occurring world-wide. These new light sources are revolutionizing both the spatial and temporal resolution of the technique, although this revolution comes in the form of data volumes that will be completely untenable for iterative phase retrieval methods, but could be handled by the developed unsupervised deep learning solution. Additionally, the unsupervised network eliminates the need for ground truth real space images in training, potentially enabling online network training during the data acquisition with minimal human intervention, which is impractical for supervised network that needs computationally intensive iterative phase retrieval to obtain ground truth images. It's worth mentioning that we note that our approach is broadly applicable to any phase retrieval problem (or more generally to inverse problems) where the forward model is accurately known. 

\section*{Results}

\subsection*{Approach}

The AutoPhaseNN framework is built by combining 3D convolutional encoder–decoder neural network with the physical model of X-ray scattering. With the loss function calculated between the measurement and the estimated diffraction generated by the forward physical model, the 3D CNN is trained to learn the inverse process of the physical model with only unlabeled simulated and/or measured diffraction patterns without needing real space images. Once trained, the physical model can be discarded and only the 3D CNN model is used to provide the direct inversion from 3D diffraction intensities to real space images without needing the iterative process. In this section, we briefly describe the neural network architecture used in the training stage. Subsequently, we demonstrate the efficacy of the trained model and the subsequent refinement process using both simulated and experimental data.

\subsection*{Unsupervised deep learning model}

The architecture of the unsupervised physics-aware deep learning model (AutoPhaseNN) is depicted in Fig. 1a. The model is based on a 3D CNN framework with a convolutional autoencoder and two deconvolutional decoders that learns a direct inversion from the diffraction intensity to the real space image amplitude and phase. In addition to the 3D CNN portion, we include the X-ray scattering model into the network architecture, including the numerical modeling of diffraction and custom layers to enforce image shape support (Fig. 1b). The outputs of the 3D CNN, amplitude and phase images, are combined to form the complex number, which is input to the X-ray scattering model to obtain the estimated reciprocal space intensity. Loss for each training example is then computed as the mean absolute error (MAE) between measured and estimated diffraction intensities. The loss function is defined as:
\begin{equation}
    Loss(I_e, I_m) = \frac{\sum{|\sqrt{I_e}-\sqrt{I_m}|}}{N^3},
\end{equation}
where $I_e$ and $I_m$ correspond to the estimated and measured diffraction intensities and $N \times N \times N$ is the size of the output image. We note that the reciprocal space intensity is also the input to the network (Fig. 1a), and hence at no stage do we directly use the ground truth of target outputs (i.e., sample amplitude and phase) to train the network.

In the 3D CNN, the encoder takes the magnitude of the 3D diffraction pattern with the size of $64\times64\times64$ as its input. Then, the encoded data is passed through two separated decoders to generate the $64\times64\times64$ amplitude and phase images in real space. The input is connected to the output using convolution blocks, max pooling, upsampling and zero padding layers. The convolution block is composed of two $3\times3\times3$ convolution layers, the leaky rectified linear unit (LRLU) activation function and the batch normalization (BN). Further details of the network architecture along with the X-ray scattering model are included in \textbf{Methods}.

Unlabeled simulated datasets generated from a physics-informed data preparation pipeline using atomistic structures and a small amount of experimental datasets acquired from BCDI experiments were used for network training and testing (see \textbf{Methods} for details of the training and testing data). Once the training is complete, we discard the X-ray scattering model and only keep the 3D CNN portion of the network which has now learnt to invert the intensity from coherent X-ray diffraction to the sample image amplitude and phase in a single step. 

\subsection*{Performance on simulated 3D CDI data}

Figure 2 shows examples of AutoPhaseNN's performance on simulated data. We tested the trained AutoPhaseNN model on $\sim$2k unseen 3D diffraction patterns that were never shown to the network during training. To evaluate the quality of the prediction, we calculated the $\chi^2$ error for the modulus of the diffraction pattern in reciprocal space defined as:
\begin{equation}
    \chi^2 = \frac{\sum{(\sqrt{I_e}-\sqrt{I_m})^2}}{\sum{I_m}},
\end{equation}
together with the structural similarity (SSIM) for the amplitude of the real space image. Figure 2 (a and b) show the distributions of the $\chi^2$ error and SSIM, indicating that the network gives excellent performance in predicting the amplitude and phase of the 3D image from the input diffraction intensity. Four representative results are displayed in Fig. 2 (c-g). The predicted 3D images (Fig. 2f) show high agreement with the corresponding ground truth images (Fig. 2d). Even for crystal 4, which has the largest computed error under the reciprocal space $\chi^2$ error metric, the network still correctly predicts the shape and phase distribution of the image. 

\subsection*{Refinement on deep learning prediction}

For comparison, we performed conventional iterative phase retrieval on the test datasets. Although the computed $\chi^2$ errors for the network prediction in Fig. 2f are relatively large compared to phase retrieval results in Fig. 2e, the network prediction could provide a learned prior to the iterative phase retrieval. The convergence of these iterative algorithms is often sensitive to initialization conditions, such as the initial guess of the object and support \cite{Marchesini2003}. Additionally, it usually requires switching algorithms, thousands of iterations, and multiple random initializations to converge to a solution with high confidence \cite{Chen2007}.  A refinement procedure was conducted on the real space images predicted by AutoPhaseNN. We performed iterative phase retrieval using the output of AutoPhaseNN model (amplitude and phase of the 3D image) as the learned prior. In Fig. 2g, we demonstrate that the reconstruction error can be further reduced with only 50 iterations of the refinement process (error reduction (ER) algorithm) and reach comparable or even lower errors compared to the phase retrieval results (which does not achieve the same level of error even after 600 iterations). 

Supplementary Figures 3 and 4 show the difference between the reconstructed and the measured diffraction data for two representative results (crystal 2 and 4), showing that the refinement process improves the reconstruction from the AutoPhaseNN prediction. Supplementary Figures 1 and 2 plot $\chi^2$ errors for 2D slices in diffraction patterns for the three approaches. It can be observed that conventional phase retrieval and the refinement have very close reconstruction errors in the low-q region, while refined results have relatively lower errors in the high-q region. This indicates that the refinement process based on the learned prior given by the network prediction has better performance in retrieving high spatial frequency information, probably because the network predictions are already very close to the true solution, especially the predicted amplitude as indicated by the SSIM value shown in Fig. 2.

\subsection*{Demonstration on experimental 3D BCDI data}

The performance of the trained AutoPhaseNN model was also evaluated on experimentally acquired data. 3D BCDI data was measured from sub-micrometer size gold crystals grown by high temperature dewetting of a thin gold film deposited on a Silicon substrate, with the native oxide intact. The crystals form with a predominately (111) crystallographic orientation normal to the substrate surface. This leads to a favored texture of (111) Bragg peaks with a random orientation about the surface normal. We isolated crystals with diffraction signals that were sufficiently isolated from the predominate (111) texture and measured three-dimensional coherent diffraction patterns in the vicinity of those individual (111) Bragg peaks. Further experimental details are described in \textbf{Methods}.

The coherent diffraction patterns from ten randomly selected crystals were used to refine the neural network that was pre-trained on the simulated data. Following a standard pre-processing step (see details in \textbf{Methods}), the performance of the refined neural network was evaluated with previously unseen experimental data. We also performed iterative phase retrieval with the same parameters as before for comparison to AutoPhaseNN's predictions. The results are shown in Fig. 3 (a and b). AutoPhaseNN provides a predicted shape and strain field of the crystal very close to the phase retrieval results with $\sim$100 times speed up, taking about 200 ms for each prediction on a CPU processor, in contrast to phase retrieval that took about 28 s for 600 iterations. 
As before, we also used AutoPhaseNN's prediction as a learned prior to iterative phase retrieval. The recovered images after just 50 iterations of phase retrieval (ER) starting from AutoPhaseNN's prediction are shown in Fig. 3c. From the visual comparison and the $\chi^2$ error, the refinement produces improved details in shape and phase and gives a comparable and even lower error compared to phase retrieval results. We note that the phase retrieval reconstruction couldn't achieve the same error even after more iterations (Supplementary Figure 6). Additionally, the combination of AutoPhaseNN prediction and the refinement procedure is still about 10 times faster than the iterative phase retrieval (600 iterations). 

Figure 4 shows the reconstruction $\chi^2$ error as a function of the iteration for the five representative results shown in Fig. 3. The dashed lines represent errors for iterative phase retrieval alone, showing three cycles of error reduction (ER) and hybrid input and output (HIO) algorithms as described in \textbf{Methods}. The $\chi^2$  for HIO rising and reaching a plateau is a sign of escaping local minima \cite{Fienup1982PhaseComparison,marchesini2007phase}, then the error decreases rapidly on iterations with the ER algorithm and becomes more consistent with the actual image quality, which causes the abruptness in the reconstruction error. The solid lines are errors for the refinement using the predicted prior, showing that the refinement procedure starts with a lower reconstruction error and converges faster compared with the conventional phase retrieval since its initial image is already very close to the true solution. Supplementary Figure 5 shows reconstruction errors for ER algorithm that starts from random initial guesses. The ER reconstructions reach similar error values as the network predictions after some numbers of iterations as indicated by the dots in Supplementary Figure 5a. But the reconstructed real space images with 500 iterations of ER algorithm (shown in Supplementary Figure 5 (b-f)) are clearly worse than both network predictions (Fig. 3b) and refined results (Fig. 3c) (50 iterations of ER initialized by network prediction), proving that the learned prior given by network prediction could improve the convergence performance of ER algorithm and relax the requirement of switching algorithms to achieve a satisfactory solution. 

In order to avoid misleading results due to over-fitting when optimizing/predicting from the diffraction data, a free R-factor was evaluated for the network prediction, conventional phase retrieval, and the refinement \cite{favre2020free}. The free Poisson log-likelihood $LLK_{free}$ and the free $\chi ^2 _{free}$ (described in SUPPLEMENTARY NOTE 3) are presented in Supplementary Table 1, which also demonstrates that the refinement process with the learned prior provided by the network gives the best performance.

\section*{Discussion}

We have demonstrated the application of the developed unsupervised physics-aware deep learning model, AutoPhaseNN, for direct inversion of the 3D BCDI data from the far-field measurement to the real space image. By incorporating the X-ray scattering model into the network architecture, AutoPhaseNN is trained with only measured diffraction patterns without needing real space images. Once trained, only the 3D CNN model is used to provide 3D images in the real space from the measured diffraction intensities in the reciprocal space. The AutoPhaseNN prediction is $\sim$100 times faster when compared to traditional iterative phase retrieval (600 iterations), which will greatly benefit real-time CDI experiments. Combined with the refinement process (50 iterations of ER algorithm), the final result is comparable to the traditional phase retrieval while being $\sim$10 times faster. We believe the AutoPhaseNN model will revolutionize the broad application of neural networks to phase retrieval problems, including CDI for which AutoPhaseNN was designed and other inverse problems in which the forward model is known.

Currently, AutoPhaseNN network is trained offline and with unlabeled simulated data and limited amount of experimental data. The current simulated data are created from face center cubic (fcc) lattice of gold with atomistic simulations. We expect the further improvement of the network performance with a large and more diverse training dataset, such as different reciprocal space oversampling ratios,crystals of different space groups and crystals with defects, etc. Practically, it is difficult to train a single network that can handle all sample types. We have demonstrated a strategy, in which the network pre-trained on simulated data is finely tuned for experimental data. This provides a practical approach to image new samples, bypassing the need to train an entirely new network from scratch. More importantly, with its ability to train without real space images, the performance of the AutoPhaseNN model can be continuously improved with data from new coherent imaging experiments. In addition, an idealized X-ray scattering model is employed in the current network training process. Further improvements can be made by optimizing the physical model of the coherent imaging, for example, including the partial coherence correction\cite{Clark2012} or dynamical scattering effects\cite{hu2018dynamic}. 

We believe this unsupervised deep learning solution for data inversion will be critical to the coherent imaging technique, especially in the light of the present and future fourth generation synchrotron sources, including the Advanced Photon Source Upgrade (APS-U) and the Extremely Brilliant Source at the European Synchrotron Research Facility (ESRF-EBS). These upgrades will provide two or three orders of magnitude increased coherent X-ray flux. With the subsequent vast increase in the coherent imaging data, traditional iterative phase retrieval methods will not be able to keep up. DL methods have been shown to produce high-fidelity images on a sub-second timescale, which can provide real-time feedback for the experiment. Additionally, unsupervised training can potentially enable online real-time improvements to the neural network with minimal human intervention, in effect creating a continuously self-improving NN capable of keeping up with the vast data rate of next generation light sources and electron microscopes. With further development, the unsupervised deep learning approach demonstrated by AutoPhaseNN, will be vital to coherent imaging experiments both now and in the future. 

\section*{Methods}

\subsection*{AutoPhaseNN architecture}
In the AutoPhaseNN model, the physical knowledge is included in the network architecture to enforce the physical constraint. As shown in Fig. 1a, two zero padding layers are employed before the last convolution blocks to act as the square support which sets the upper size limit of the real space object. The square support is set to half the size of the input diffraction data so that the oversampling condition for phase retrieval is met \cite{miao2000oversampling}. In addition, the physical knowledge that the normalized amplitude is within the interval [0,1] and the phase falls within the interval [-$\pi$,+$\pi$] is built into the network architecture using the Sigmoid activation function and the Tanh activation function in the final 3D convolution layers in the amplitude and phase branches, respectively. 

Figure 1b describes the detailed X-ray scattering forward model during the training stage, including the numerical modeling of diffraction and image shape constraints. The predicted amplitude and phase are combined to generate the complex image. An image shape support function is obtained by thresholding the current predicted amplitude using a contour at the 10$\%$ intensity level. This image shape support is continuously updating from the predicted amplitude and is used only once per image per epoch during the network training. The aforementioned square support enforced by the zero padding layers and this image shape support can impose additional constraints during the unsupervised training process. The physical constraints implemented in the network architecture and the scattering forward model provide strong prior information that assures the performance of the unsupervised model without the labeled data.

The estimated diffraction pattern is obtained from the FT of the current estimation of the real space image that generated from the network output. The network weight and bias factors are optimized with the objective function that minimizes the loss between the input diffraction pattern and the estimated diffraction pattern. By incorporating the physics, the ground truth of the real space image is not needed during training. Once a trained model has been obtained, only the 3D CNN model is kept to recover the amplitude and phase information of the measured sample. 

\subsection*{Simulated data generation}

Every crystal in the training dataset is prepared from a $\approx$ $20$ nm $\times$ $20$ nm $\times$ $20$ nm cube of face center cubic (fcc) lattice of gold. To create diverse shape and size, each crystal has $4$ -- $20$ facets generated by clipping along randomly chosen high crystal symmetry planes that are positioned at random distances from the geometric center of the crystal. To create realistic local strains within the crystal, a combination of compression, tension, and shear stresses (up to 1\% strain) is applied to the initial atomistic crystal structure. Subsequently, Large-scale Atomic/Molecular Massively Parallel Simulator (LAMMPS) is used to relax the strained structure via energy minimization performed using the embedded-atom method (EAM) interatomic potential for gold. After minimization, the lattice constant of the initial and final structures are normalized to 1 (i.e., scaled by the inverse lattice constant of gold, 1/4.078 \AA$\;$), and a $\approx$ 5 lattice unit padding ($\approx$ 20 \AA$\;$ before lattice normalization) is added to each side of the normalized structures to avoid any potential boundary related artifacts. Finally, the output object is created from these structures, which corresponds to a complex crystal density composed of the crystal shape (i.e., number density of atoms) and its local strain (i.e., lattice phases). The number density of atoms and lattice phases are computed using an uniform grid with a bin size of $\approx$ 2 lattice units ($\approx$ 8 \AA$\;$ before lattice normalization). The number density values are normalized by the maximum density value whereas the lattice phases are computed from the atom displacement field projected along [111] and are scaled by 2$\pi$. This binning process converts the crystal atomistic model to a $32\times32\times32$-sized object. The complex object array is then zero padded to twice the size ($64\times64\times64$). The 3D diffraction pattern is generated via FT of the complex object and Poisson distributed noise is added. The magnitude of the diffraction pattern is the input for the network and is the only thing needed in training. We note that although simulated crystals generated using this method are smaller than experimentally measured particles, the approach is still valid considering that the CNN is pixel-based and scale-independent.

\subsection*{3D CDI data acquisition}

The standard BCDI scan was a one hundred point rocking curve of +/- 0.5 degrees of the omega axis of the six circle diffractometer\cite{lohmeier1993angle}. All experimental data was acquired at the 34-ID-C beamline of the Advanced Photon Source at Argonne National Laboratory. The X-ray beam was set to 9 keV and focused using Kirk-Patrick Baez mirrors to approximately 700$\times$700 $\textrm{nm}^2$. A Medipix2 (Timepix) detector was positioned at either 500 mm or 750 mm from the sample depending on the size of the crystal.  The detector distance was determined by the need to oversample the diffraction intensities with the 55 ${\mu}$m pixel size of the Timepix detector.

\subsection*{AutoPhaseNN training}
The AutoPhaseNN model was implemented based on the Keras package running the Tensorflow backend. The training dataset is a combination of simulated diffraction patterns and experimental diffraction patterns.As described in \textbf{Simulated data generation}, simulated 3D crystals were derived from a physics-informed data preparation pipeline, creating a wide variety of amplitude and phase states. The unlabeled simulated data consists of 54,028 3D diffraction patterns generated from different crystal shapes, within which 52k data are used for training and 2028 are reserved for testing. A rotation data augmentation was performed on the 52k simulated diffraction patterns to generate 104k data for training. The entire training dataset was randomly split, giving 80$\%$ used for training and the remaining 20$\%$ reserved for validation. When training the model, adaptive moment estimation (ADAM) with an initial learning rate of 0.001 is used to update the weights and biases of the model. At the end of each epoch, the performance of the network was evaluated using the validation data. The ReduceLROnPlateau callback was used to reduce the learning rate by a factor of 2 when the validation loss has stopped improving. The network was trained in parallel on 8 NVIDIA A100 GPUs for 70 epochs. The training took $\sim$12 h with a mini-batch size of 16 on each GPU. 

Before applying AutoPhaseNN on experimental data, the network weights were fine-tuned with an augmented dataset generated from a small amount of experimental data. A data augmentation process consisting of rotation, resampling, and flipping was performed on 10 diffraction patterns acquired from BCDI experiments to generate 3200 experimental training data. We took the weights from the network pre-trained with simulated data and continued the training with the experimental training data (as presented in Supplementary Figure 7). The fine-tuning took $\sim$ 20 min on 8 GPUs for 50 epochs with an initial learning rate of $10^{-6}$ and a mini-batch size of 8 on each GPU.

\subsection*{Pre-processing experimental data}
AutoPhaseNN was mostly trained on unlabeled simulated data with the reciprocal space oversampling ratio of around 3.0. When tested on experimental data, we prepared input data by downsampling the 3D coherent diffraction pattern of the gold crystal acquired from BCDI experiments to yield an oversampling ratio of about 3.0. The $64\times64\times64$ sized input data was then obtained by cropping the downsampled 3D diffraction pattern. 

\subsection*{Phase retrieval method}
For conventional iterative phase retrieval, the $64\times64\times64$ diffraction pattern was input to iterative phase retrieval, where the algorithm was switched between error reduction (ER) and hybrid input-output (HIO). The algorithm was performed with the combination of 20 iterations of ER, 160 iterations of HIO, and 20 iterations of ER as one cycle. Three cycles (600 iterations) were performed in total using shrink-wrap support in real space.

For the refinement procedure, the predicted amplitude and phase obtained from the AutoPhaseNN model were provided as the initial image guess and the initial support was generated by thresholding the predicted amplitude. Then 50 iterations of ER algorithm were performed to generate the final refined result. 

\section*{Acknowledgments}
This work was performed, in part, at the Advanced Photon Source, a U.S. Department of Energy (DOE) Office of Science User Facility, operated for the DOE Office of Science by Argonne National Laboratory under Contract No. DE-AC02-06CH11357. This research used resources of the Argonne Leadership Computing Facility, which is a DOE Office of Science User Facility supported under Contract DE-AC02-06CH11357. We gratefully acknowledge the computing resources provided on Swing, a high-performance computing cluster operated by the Laboratory Computing Resource Center at Argonne National Laboratory. This work was supported by the U.S. Department of Energy, Office of Science, Office of Basic Energy Sciences Data, Artificial Intelligence and Machine Learning at DOE Scientific User Facilities program under Award Number 34532. M.J.C. acknowledges partial support from Argonne LDRD 2021-0090 – AutoPtycho: Autonomous, Sparse-sampled Ptychographic Imaging. Y.Y. acknowledges partial support from Argonne LDRD 2021-0315 – Scalable DL-based 3D X-ray nanoscale imaging enabled by AI accelerators.
\section*{Author contributions} Y.Y. and M.J.C. proposed the initial idea, R.J.H. contributed to the conceptualization. H.C. generated the simulated data. R.J.H. collected the BCDI experimental data. Y.Y. built the model and performed network training and testing with the help from M.J.C., R.J.H. and H.C.. P.B. gave technical support and conceptual advice. M.J.C., Y.Y., R.J.H. and H.C. wrote the manuscript with input from P.B. and S.S..
\section*{Competing interests}Authors declare that they have no competing interests.
\section*{Data and materials availability} All data needed to evaluate the conclusions in the paper are present in the paper. The codes and trained models developed in this study are available in a public GitHub repository at \url{ https://github.com/YudongYao/AutoPhaseNN}. 




\begin{figure}[p]
\includegraphics[width=\textwidth]{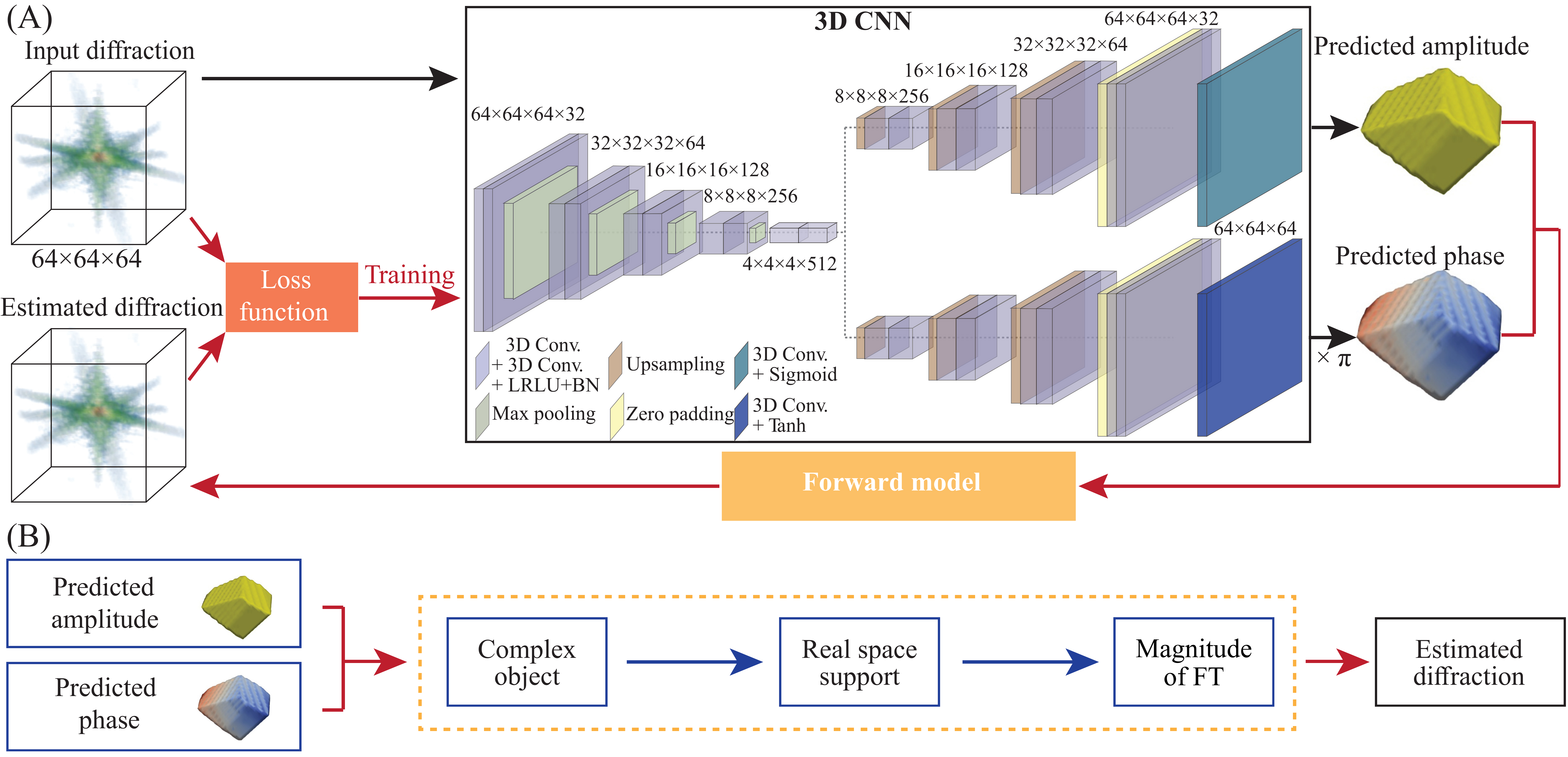}
\caption{\label{fig:one} \textbf{Schematic of the neural network structure of AutoPhaseNN model during training.} (\textbf{a}) The model consists of a 3D CNN and the X-ray scattering forward model. The 3D CNN is implemented with a convolutional auto-encoder and two deconvolutional decoders using the convolutional, maximum pooling, upsampling and zero padding layers. The physical knowledge is enforced via the Sigmoid and Tanh activation functions in the final layers. (\textbf{b}) The X-ray scattering forward model includes the numerical modeling of diffraction and the image shape constraints. It takes the amplitude and phase from the 3D CNN output to form the complex image. Then the estimated diffraction pattern is obtained from the FT of the current estimation of the real space image.}
\end{figure}

\begin{figure}[ht]
\includegraphics[width=\textwidth]{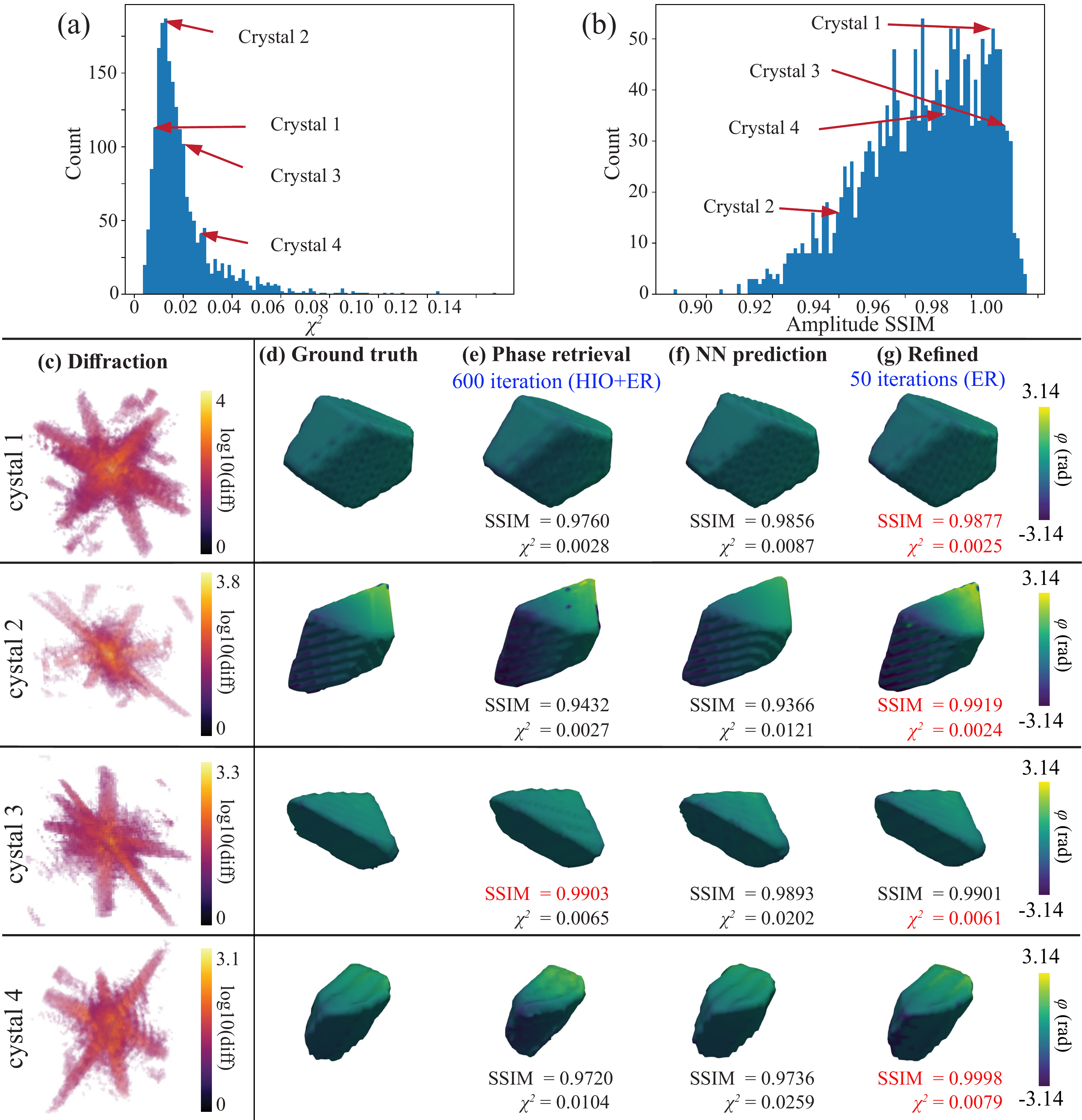}
\caption{\label{fig:two} \textbf{Performance of AutoPhaseNN on simulated test data.} (\textbf{a}) Histogram of $\chi^2$ error for the modulus of the diffraction pattern. (\textbf{b}) Histogram of SSIM for the amplitude of the real space images. (\textbf{c}) Volume rendering of the input 3D diffraction patterns, (\textbf{d}) ground truth images, (\textbf{e}) phase retrieval results, (\textbf{f}) network predictions, and (\textbf{g}) refined results, for the four representative samples. (\textbf{d-g}) show the 0.3 contour of the amplitude and the color represents the phase on the surfaces. Reciprocal $\chi^2$ errors and SSIMs for the amplitude of real space images are shown in the figures.The best figure of merit for each crystal is marked in red.}
\end{figure}

\begin{figure}[ht]
\includegraphics[width=\textwidth]{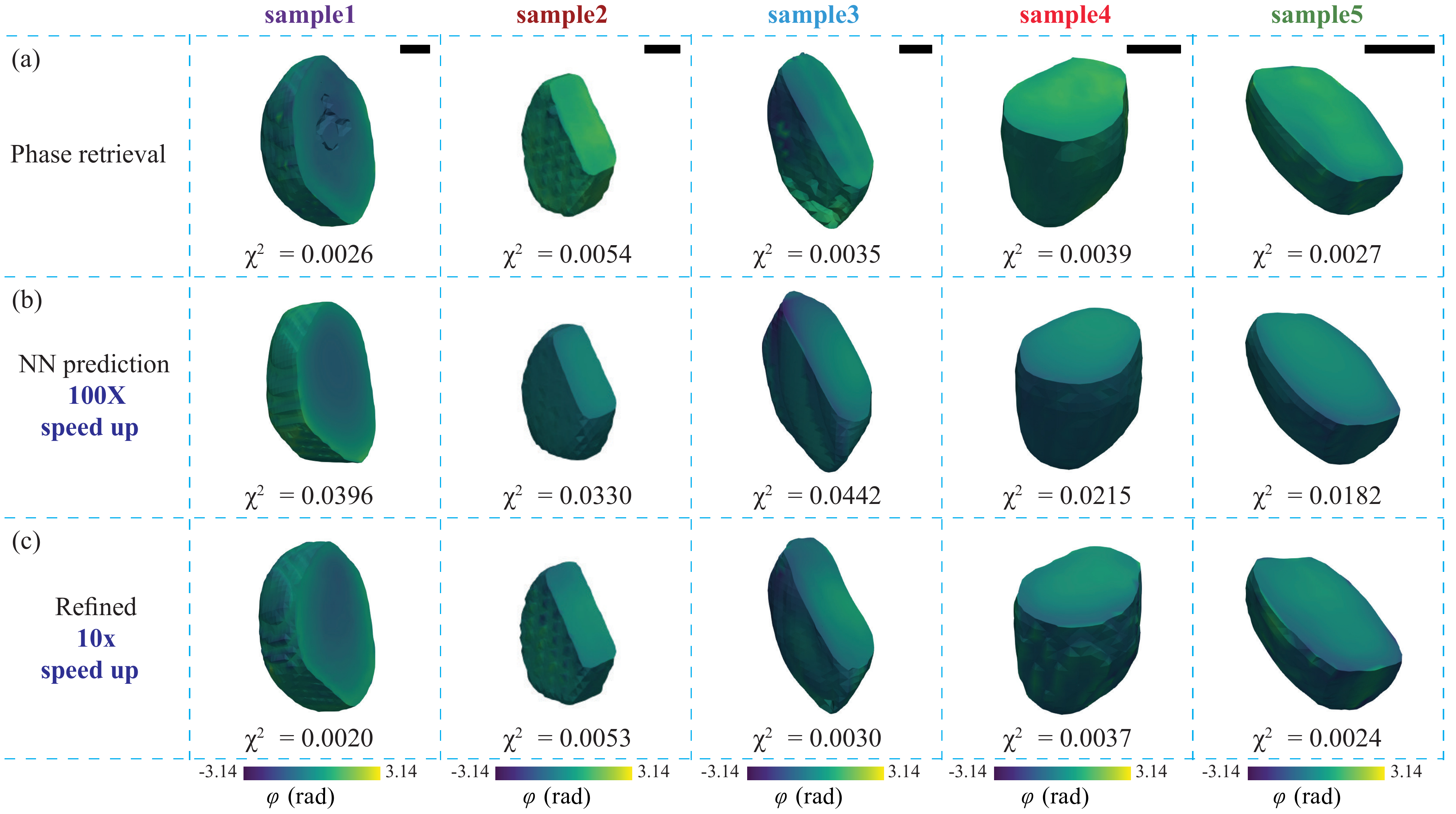}
\caption{\label{fig:three} \textbf{Performance of AutoPhaseNN on real data from BCDI experiments.} Comparison among reconstructions from (\textbf{a}) traditional phase retrieval, (\textbf{b}) prediction of the AutoPhaseNN model, and (\textbf{c}) refinement process. The crystals are clipped to show the internal strain fields and the images show the 0.3 contour of the amplitude and the color represents the phase. Reciprocal $\chi^2$ errors are shown in the figures. The scale bars are all 150 nm.}
\end{figure}

\begin{figure}[ht]
\includegraphics[width=\textwidth]{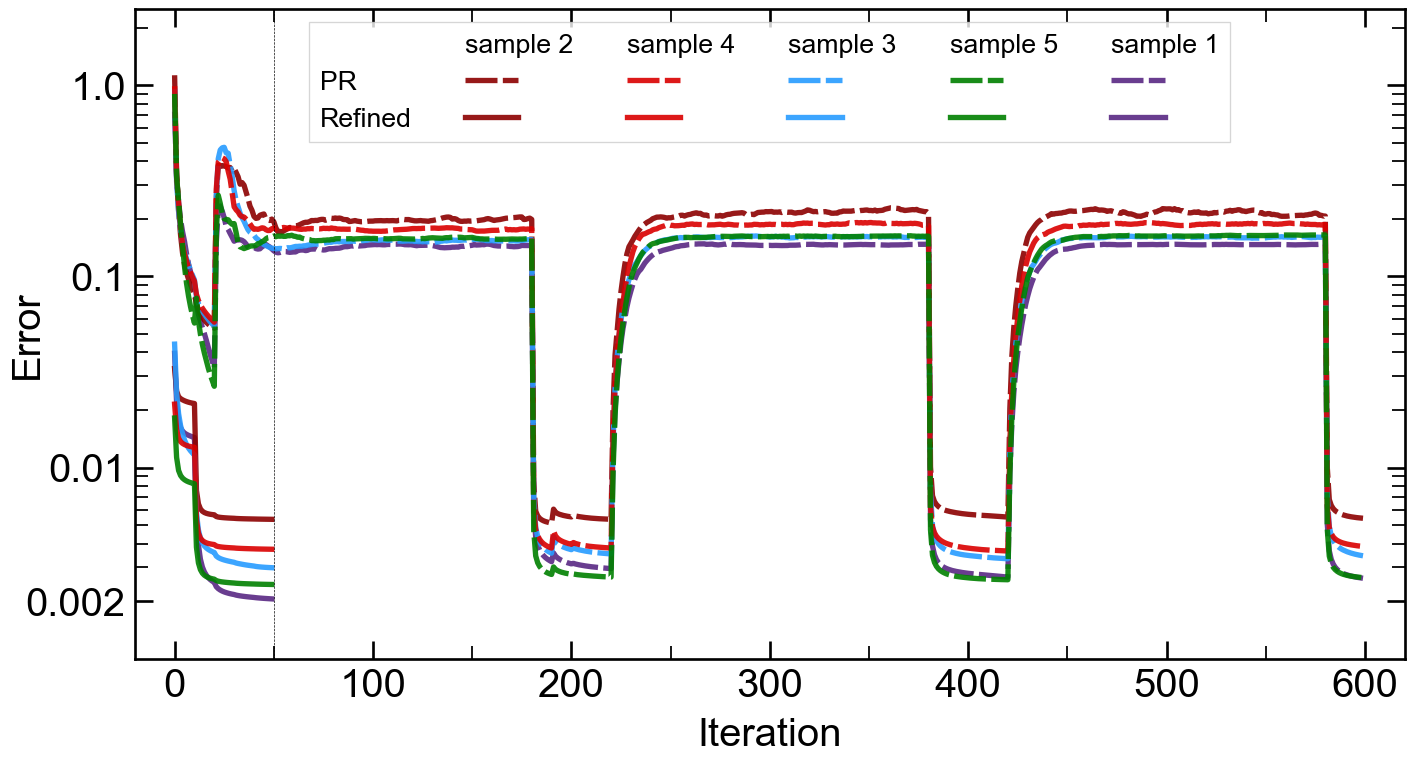}
\caption{\label{fig:four} \textbf{Reconstruction error as a function of the iteration number for conventional phase retrieval and refinement.} The dashed lines represent the error for conventional phase retrieval while the solid lines are for refinement process.}
\end{figure}

\end{document}